\documentclass[a4paper,groupaddress]{jpconf}
\usepackage{graphicx}
\begin{document}
\title{Project 8 Phase III Design Concept}


\author{
A~Ashtari~Esfahani$^1$,
S~B{\"o}ser$^2$,
C~Claessens$^2$,
L~de Viveiros$^3$,
P~J~Doe$^1$,
S~Doeleman$^4$,
M~Fertl$^1$,
E~C~Finn$^5$,
J~A~Formaggio$^6$,
M~Guigue$^5$,
K~M~Heeger$^7$,
A~M~Jones$^5$,
K~Kazkaz$^8$,
B~H~LaRoque$^3$,
E~Machado$^1$,
B~Monreal$^3$,
J~A~Nikkel$^7$,
N~S~Oblath$^5$,
R~G~H~Robertson$^1$,
L~J~Rosenberg$^1$,
G~Rybka$^1$,
L~Salda{\~n}a$^7$,
P~L~Slocum$^7$,
J~R~Tedeschi$^5$,
T~Th{\"u}mmler$^9$,
B~A~VanDevender$^5$,
M~Wachtendonk$^1$,
J~Weintroub$^4$,
A~Young$^4$ and
E~Zayas$^6$
}

\address{
$^1$ Center for Experimental Nuclear Physics and Astrophysics, and Dpt. of Physics, University of Washington,        Seattle, WA, USA\\
$^2$ Johannes Guttenberg University, Mainz, Germany\\
$^3$ Dept. of Physics, University of California, Santa Barbara, CA, USA\\
$^4$ Harvard-Smithsonian Center for Astrophysics, Cambridge, MA, USA\\
$^5$ Pacific Northwest National Laboratory, Richland, WA, USA\\
$^6$ Laboratory for Nuclear Science, Massachusetts Institute of Technology, Cambridge, MA, USA\\
$^7$ Dept. of Physics, Yale University, New Haven, CT, USA\\
$^8$ Lawrence Livermore National Laboratory, Livermore, CA, USA\\
$^9$ Karlsruhe Institute for Technology, Karlsruhe, Germany
}
\ead{mathieu.guigue@pnnl.gov}

\begin{abstract}
We present a working concept for Phase III of the Project 8 experiment, aiming to achieve a neutrino mass sensitivity of $2~\mathrm{eV}$ ($90~\%$ C.L.) using a large volume of molecular tritium and a phased antenna array.
The detection system is discussed in  detail.
\end{abstract}

\section{Project 8  and Cyclotron Radiation Emission Spectroscopy}

The Project 8 Collaboration is taking a four-phase approach to implement a next-generation tritium endpoint experiment to measure the neutrino mass. 
Based on the technique of Cyclotron Radiation Emission Spectroscopy (CRES) \cite{Monreal2009}, we construct an energy spectrum from the cyclotron frequencies of magnetically trapped electrons. 
Phase I, now completed, has shown the feasibility of this technique using $^{83\mathrm{m}}\mathrm{Kr}$ \cite{Asner2015}.
Phase II, which uses a cylindrical waveguide and a cell compatible with tritium, is underway. 
By increasing the volume by more than one order of magnitude, Phase III targets a neutrino mass sensitivity comparable to current limits from the Mainz and Troitsk experiments \cite{Kraus2005,Aseev2011}: $m_{\beta}\leq 2~\mathrm{eV}$ \cite{Olive2014}.
Using an atomic tritium source and a even larger fiducial volume, Phase IV aims at a 40-meV sensitvity to the neutrino mass.

\section{Phased Array of Antennas}

\begin{figure}
\begin{center}
\includegraphics[width=0.3\textwidth]{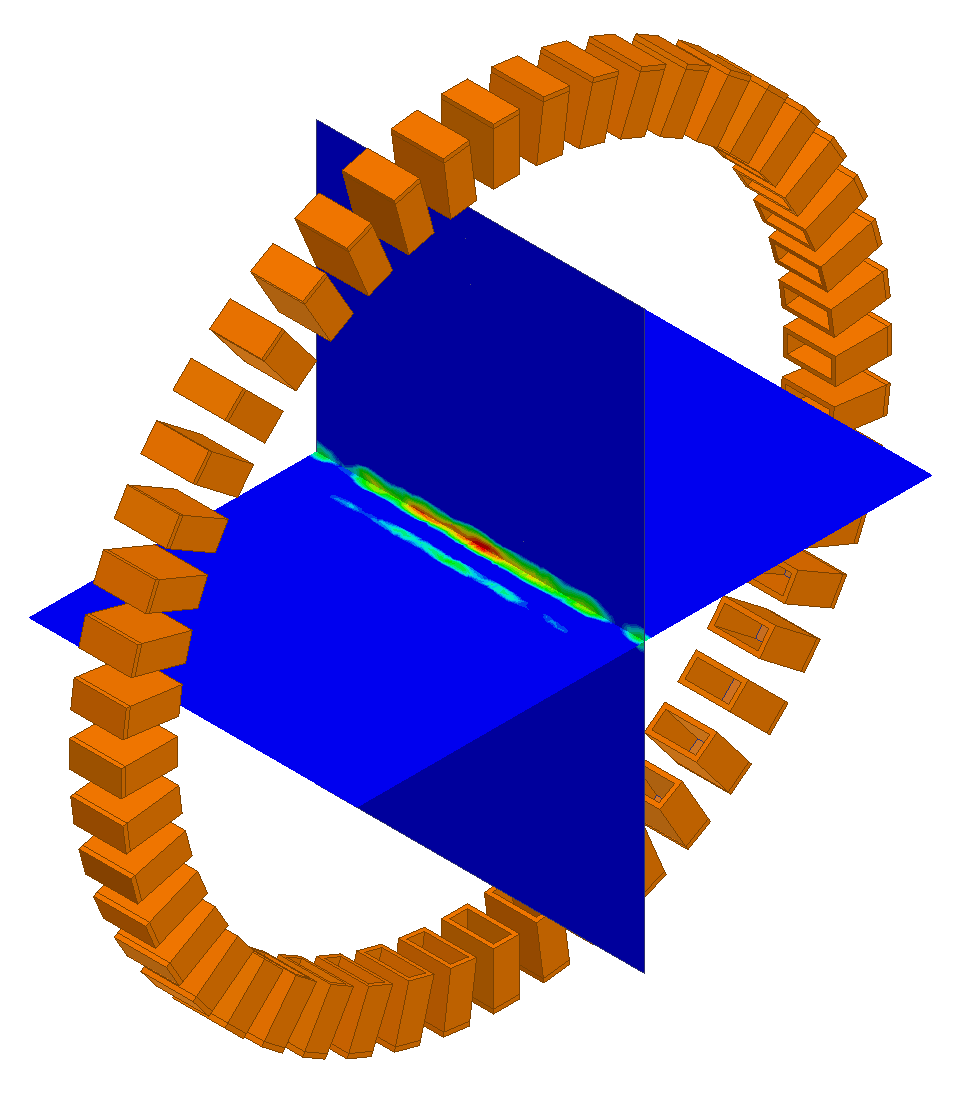}
\includegraphics[width=0.65\textwidth]{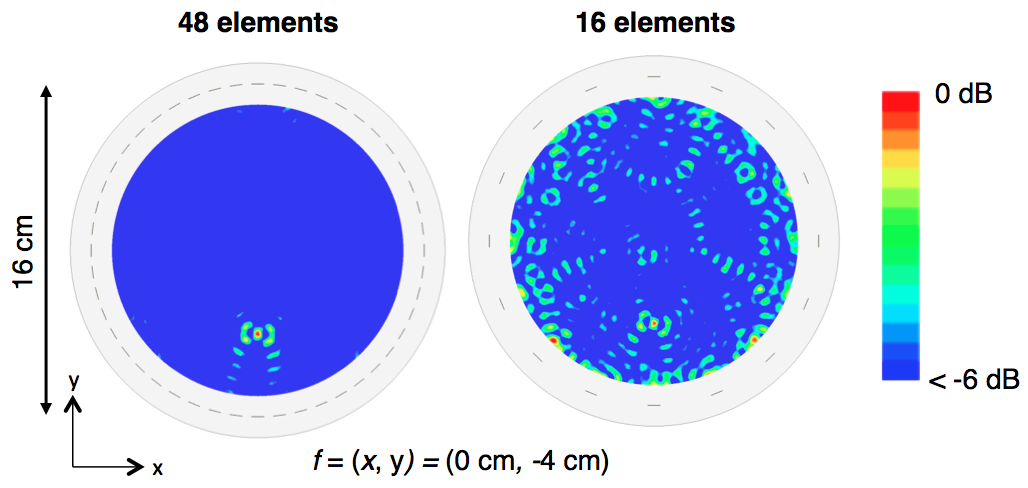}
\caption{\label{fig:array} Left: Possible 48-element ring array of antennas for Phase III. 
Colors above blue represents the -6-dB focal region of the phased array.
Right: Gain patterns for 8-cm radius ring arrays with 48- and 16-element phase-focused $4~\mathrm{cm}$ from the center. 
A 48-element array posseses a smaller number of grating lobes compared with the 16-element one, which makes the electron position reconstruction more accurate.}
\end{center}
\end{figure}

One can achieve Phase III sensitivity after one year of data taking with a $200~\mathrm{cm}^3$ volume of tritium gas at a density of $3\times 10^{12}$ molecules per $\mathrm{cm}^3$ and a trapping efficiency of $5~\%$. 
In such an increased volume, the cyclotron radiation of beta decay electrons is essentially emitted into free space. 
The radiation can be collected using a ring-shaped array of antennas. 
Figure \ref{fig:array} shows an $8~\mathrm{cm}$ radius ring array of 48 antennas.
By appropriately phasing the antennas, one can focus the signal detection to a $1~\mathrm{mm}$ diameter region of the detector.

\begin{figure}
\begin{center}
\includegraphics[width=0.55\textwidth]{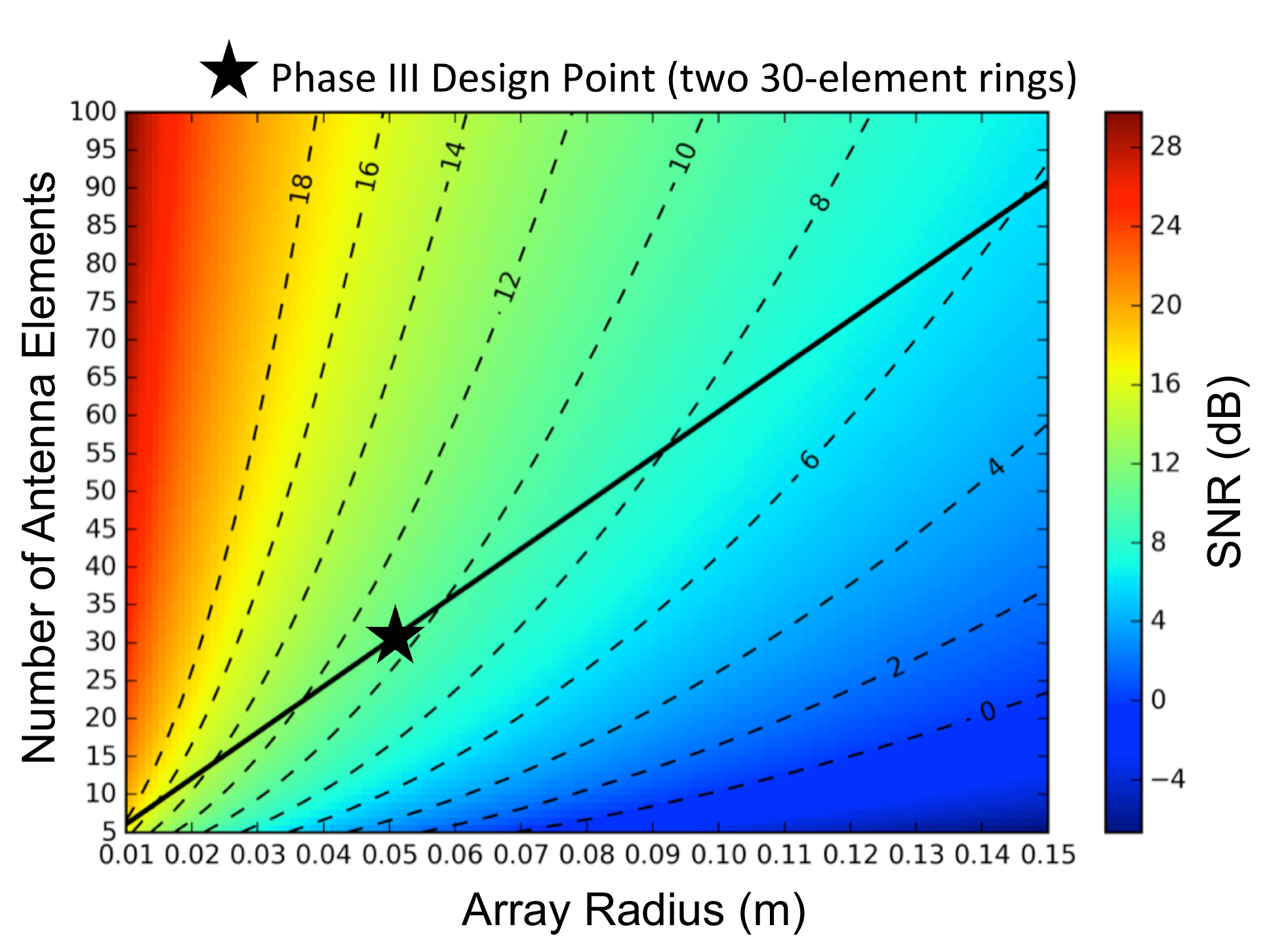}
\caption{\label{fig:noise} Signal to noise ratio as a function of the antenna ring radius and the number of elements in the array. The black line represents the maximum number of elements which can fit on the ring as a function of its radius. The black star represents the Phase III design point: two 30-elements rings.}
\end{center}
\end{figure}

A focused array can efficiently collect the power of a coherent signal while rejecting the incoherent noise.
Figure \ref{fig:noise} presents the expected SNR as a function of the array radius and the number of antenna elements in the array.
One 5-cm radius ring of 30 elements can provide a 10-dB SNR, which is sufficient for triggering purposes.

If the number of elements in a ring is not sufficient, \textit{grating lobes} appear.
Figure \ref{fig:array} shows the gain pattern for a 48-element ring and for a 16-element ring.
For the 16-element ring, the focusing point is not unique, and one cannot determine the exact position of a detected electron.
If the position of the detected electron and thus the magnetic field experienced by this electron are not known precisely, systematic uncertainties will be induced when converting the measured frequency into an electron energy.

In order to increase the instrumented volume, we propose to use two 5-cm radius rings of 30 elements each, separated by $4$-$5~\mathrm{cm}$ which can provide a 10-dB SNR. 
In this configuration the number of grating lobes is reduced.

\section{Digital Beam Forming}

\begin{figure}
\begin{center}
\includegraphics[width=0.7\textwidth]{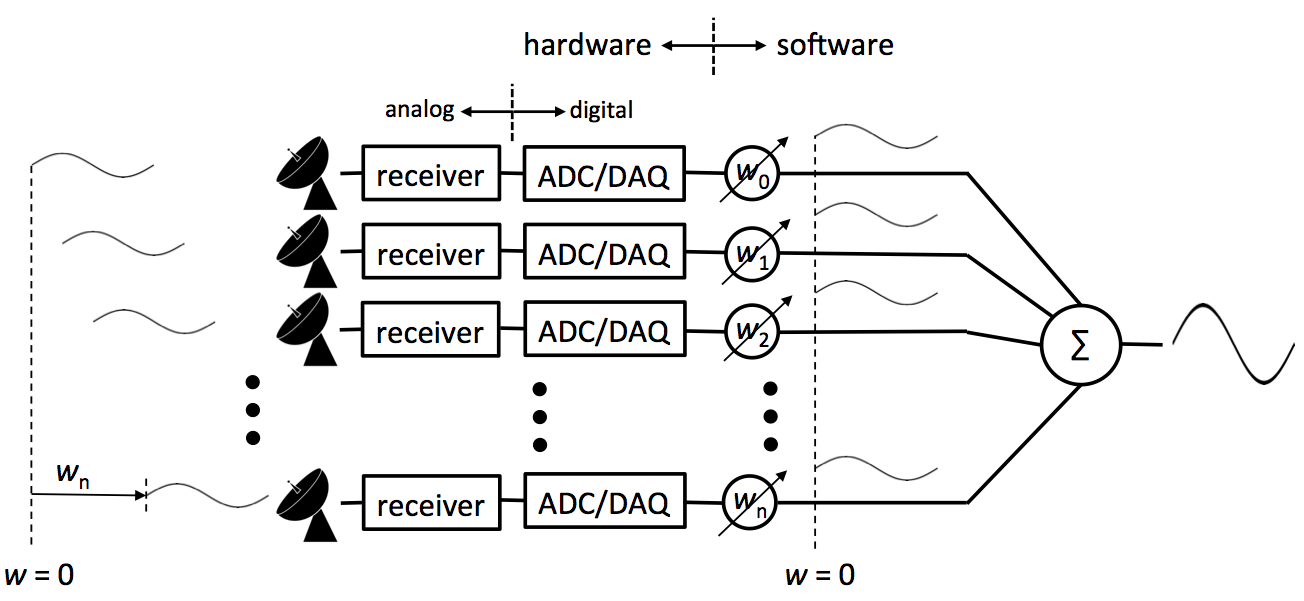}
\caption{\label{fig:digital-beam-forming} Schematic of the DAQ system for Phase III (see text for details). The $w_i$ represent the digital phase shifting and the $\Sigma$ the summation. }
\end{center}
\end{figure}

Due to the small size of the focusing point, the antennas cannot have a fixed focus.
Instead the Project 8 experiment will use a digital beam forming, similar to the procedure used by radio telescope arrays, in order to achieve a reduction of the number of channels that record the signal.
Figure \ref{fig:digital-beam-forming} presents the DAQ system for Phase III.
The signal is continuously measured by each antenna of the two-ring array.
Each antenna has a receiver that amplifies the signal via a 30-K low-noise amplifier and down-mixes the $26~\mathrm{GHz}$ signal.
Each signal is then digitalized using a ROACH2 DAQ system.
The 60 signals are sent to a GPU computing server which will handle the digital phase shifting and the channel summation.
Detection of excess power in recombined signal is used for triggering, and the summed signal is recorded  for event reconstruction on a computing cluster.

\section{Summary and Outlook}

We presented the working concept of Phase III.
A volume of $200~\mathrm{cm}^3$ is instrumented with two 5-cm rings of antennas, separated by $4$-$5~\mathrm{cm}$.
Assuming a trapping efficiency of $5~\%$ will allow us to achieve a $10~\mathrm{cm}^3$ effective volume required for a constraint on the electron neutrino mass of $2~\mathrm{eV}$ ($90~\%$ C.L.).
The magnetic trap geometry, the hardware triggering system and the reconstruction algorithm for digital beam forming are currently under development.

\end{document}